# Run-time reconfigurable multi-precision floating point multiplier design for high speed, low-power applications

Arish S
School of VLSI design and Embedded Systems
National Institute of Technology Kurukshetra
Kurukshetra, India
arishsu@gmail.com

R.K.Sharma
School of VLSI Design and Embedded Systems
National Institute of Technology Kurukshetra
Kurukshetra, India
rksharama@nitkkr.ac.in

*Abstract*: *Floating point multiplication is one of the crucial operations in many application domains such as image processing, signal processing etc. But every application requires different working features. Some need high precision, some need low power consumption, low latency etc. But IEEE-754 format is not really flexible for these specifications and also design is complex. Optimal run-time reconfigurable hardware implementations may need the use of custom floating-point formats that do not necessarily follow IEEE specified sizes. In this paper, we present a run-time-reconfigurable floating point multiplier implemented on FPGA with custom floating point format for different applications. This floating point multiplier can have 6 modes of operations depending on the accuracy or application requirement. With the use of optimal design with custom IPs (*Intellectual Properties*), a better implementation is done by truncating the inputs before multiplication. And a combination of Karatsuba algorithm and Urdhva-Tiryagbhyam algorithm (Vedic Mathematics) is used to implement unsigned binary multiplier. This further increases the efficiency of the multiplier.*

I. INTRODUCTION

Floating point multiplication units are essential Intellectual Properties (IP) for modern multimedia and high performance computing such as graphics acceleration, signal processing, image processing etc. There are lot of effort is made over the past few decades to improve performance of floating point computations. Floating point units are not only complex, but also require more area and hence more power consuming as compared to fixed point multipliers. And the complexity of the floating point unit increases as accuracy becomes a major issue. Even a minute error in accuracy can cause major consequences. These errors are possible in floating point units mainly because of the discrete behavior of the IEEE-754 [1] floating point representation, where fixed number of bits is used to represent numbers. Due to the high computational requirements of scientific applications such as computational geometry, climate modeling, computational physics, etc., it is necessary to have extreme precision in floating point calculations. And these increased precision may not be provided with single precision or double precision format. That further increases the complexity of the unit. But some applications do not require high precision. Even an approximate value will be sufficient for the correct operation. For applications which require lower precision, the use of double precision or quadruple precision floating point units will be a luxury. It wastes area, power and also increases latency.

For devices such as portable or wearable devices in which accuracy requirement varies with different applications and also power consumption is a very important factor, use of high precision floating point multipliers is not a good option. In such cases a variable precision multiplier will be a good option which can save much power and time when application doesn't need high precision. There are a lot of such models like [2], [3] and [4]. Most of such designs make use of already available IPs such as DSP (Digital Signal Processing) units and 18x18 multiplier units. In this proposed paper, we present a power efficient design of floating point multiplier with different modes of accuracy selection. With different precision modes, we can select the mode which is appropriate for the concerned application. As accuracy requirement decreases, the width of multiplier decreases and hence the power consumption and latency.

II. PROPOSED MODEL

The proposed model is a reconfigurable multi-precision floating point multiplier which can be operated in six different modes according to the accuracy requirements. It can perform floating point format multiplication of different mantissa sizes depending on the precision requirement. The basic unit is a Double-precision floating point unit. According to the precision selected, the size of the mantissa is varied. Fig. 1 shows the floating-point multiplication format used in the proposed model.

The multiplier accepts two inputs each of 67-bit wide. The first 3 bits are used for mode selection. The inputs to the multiplier can be given in double-precision floating point format with first 3 bits ($66^{th}$ bit to $64^{th}$ bit) as mode select bits.

| 3 | 1 | 11 | 52 |
|---|---|---|---|
| Mode select | Sign | Exponent | Mantissa |

Fig. 1 Floating point format used in the proposed model

The value of the mode select bits for both the inputs must be the same, otherwise a mode select error signal will be generated and the execution will be stopped. The different mode select bit combinations for different modes is shown in table 1.

TABLE I - Different modes

| Mode | Mode select bits |
|---|---|
| Mode 1(Auto Mode) | 000 |
| Mode 2 | 001 |
| Mode 3 | 010 |
| Mode 4 | 011 |
| Mode 5 | 100 |
| Mode 6 | 101 |

The different modes in the proposed multi-precision multiplier are the following.

Mode 1: Mode 1 is auto mode, i.e. the controller itself will select the optimum mode by analyzing the inputs and will start execution. The optimum mode is selected by counting the number of zeroes after a leading 1. If the number of zeroes is 6 or more after a leading 1, then the bits up to that leading 1 is counted. If the number of bits up to that leading 1 is less than 8, then mode 2 or 8-bit mantissa mode will be selected. If the number of bits before the leading 1 is less than 16, 16-bit mantissa mode will be selected and so on.

Mode 2: This is a custom precision format. It uses a basic double-precision floating point multiplier with a mantissa size of 8-bit.

Mode 3: This is a custom precision format. It uses a basic double-precision floating point multiplier with a mantissa size of 16-bit.

Mode 4: This is a custom precision format. It uses a basic double-precision floating point multiplier with a mantissa size of 23-bit.

Mode 5: This is a custom precision format. It uses a basic double-precision floating point multiplier with a mantissa size of 36-bit.

Mode 6: This mode is a fully fledged double-precision floating point multiplier at the cost of accuracy.

The modes with less number of mantissa bits consumes less amount of power. These modes are best suited for integer multiplication and also for applications where accuracy is not a big issue. Rounding of bits is done before multiplication for every mode except mode 6 and this reduces huge variations in results.

A simple block diagram of the proposed model is shown in fig. 2. The custom precision formats with 8-bit mantissa and 16-bit mantissa are best suited for integer multiplications where fractional accuracy is not an issue. It can also be used for low value fractional multiplication which require an integer value as result. By using 8-bit and 16-bit multipliers instead of a fully-fledged double precision floating point multiplier can save a lot of power and can increase the speed. The binary unsigned multiplier used for mantissa multiplication is implemented by using a combination of Karatsuba algorithm [4, 5] and Urdhva-Tiryagbhyam [6] algorithm, which gives better optimization in terms of speed

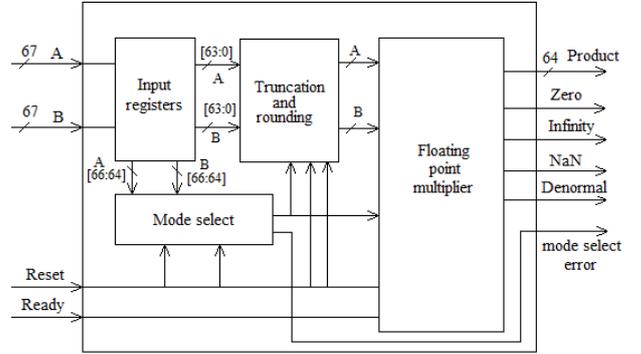

Fig. 2 Block diagram of the proposed model

and area.

### III. FLOATING POINT MULTIPLIER

A floating point number is represented in IEEE-754 format [1] as $\pm s \times b^e$ or $\pm significand \times base^{exponent}$ [7]. To perform multiplication of two floating point numbers $\pm s1 \times b^{e1}$ and $\pm s2 \times b^{e2}$, the significant or mantissa parts are multiplied to get the product mantissa and exponents are added to get the product exponent. i.e.; the product is $\pm(s1 \times s2) \times b^{(e1+e2)}$. The hardware block diagram of floating point multiplier is shown in fig. 3.

The important blocks in the implementation of proposed floating point multiplier is described below [8].

*A. Sign Calculation*

The MSB of floating point number represents the sign bit. The sign of the product will be positive if both the numbers are of same sign and will be negative if numbers are of opposite sign. So, to obtain the sign of the product, we can use a simple XOR gate as the sign calculator.

*B. Addition of Exponents*

To get the product exponent, the input exponents are added together. Since we use a bias in the floating point format exponent, we need to subtract the bias from the sum of exponents to get the actual exponent. The value of bias is $127_{10}$ ($01111111_2$) for single precision format and $1023_{10}$($01111111111_2$) for double precision format. In proposed custom precision format also, a bias of 127 is used. The computational time of mantissa multiplication operation is much more than the exponent addition. So a simple ripple carry adder and ripple borrow subtracter is optimal for exponent addition.

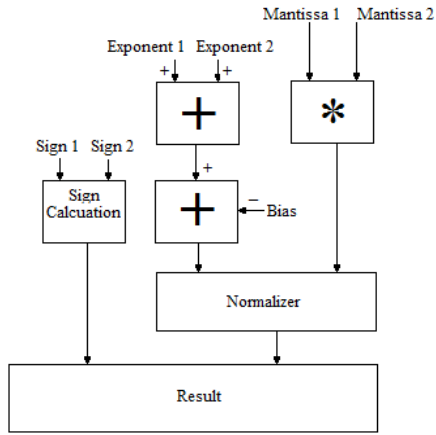

Fig. 3 Floating point multiplier

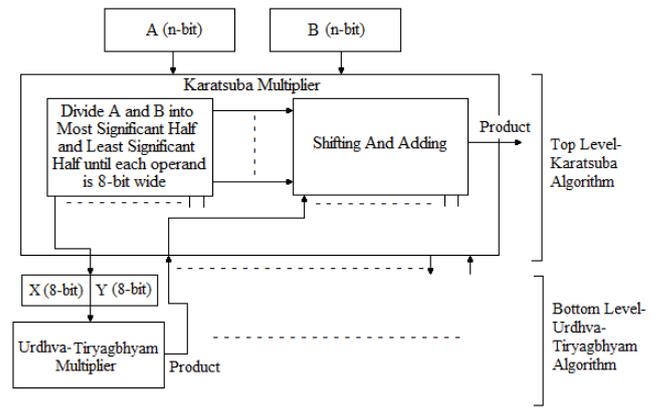

Fig. 4 Karatsuba-Urdhva multiplier model

*C. Karatsuba-Urdhva Tiryagbhyam binary multiplier*

In floating point multiplication, most important and complex part is the mantissa multiplication. Multiplication operation requires more time compared to addition. And as the number of bits increase, it consumes more area and time. In double precision format, we need a 53x53 bit multiplier and in single precision format we need 24x24 bit multiplier. It requires much time to perform these operations and it is the major contributor to the delay of the floating point multiplier. To make the multiplication operation more area efficient and faster, the proposed model uses a combination of Karatsuba algorithm and Urdhva Tiryagbhyam algorithm.

Karatsuba algorithm uses a divide and conquer approach where it breaks down the inputs into Most Significant half and Least Significant half and this process continues until the operands are of 8-bits wide. Karatsuba algorithm is best suited for operands of higher bit length. But at lower bit lengths, it is not as efficient as it is at higher bit lengths. To eliminate this problem, Urdhva Tiryagbhyam algorithm is used at the lower stages. The model of Urdhva-Tiryagbhyam algorithm is shown in Fig. 4. Urdhva Tiryagbhyam algorithm is the best algorithm for binary multiplication in terms of area and delay. But as the number of bits increases, delay also increases as the partial products are added in a ripple manner. For example, for 4-bit multiplication, it requires 6 adders connected in a ripple manner. And 8-bit multiplication requires 14 adders and so on. Compensating the delay will cause increase in area. So Urdhva Tiryagbhyam algorithm is not that optimal if the number of bits is much more. If we use Karatsuba algorithm at higher stages and Urdhva Tiryagbhyam algorithm at lower stages, it can somewhat compensate the limitations in both the algorithms and hence the multiplier becomes more efficient. The circuit is further optimized by using carry select and carry save adders instead of ripple carry adders. This reduces the delay to a great extent with minimal increase in hardware. These two algorithms are explained in detail in the below sections.

*Urdhva Tiryagbhyam algorithm for multiplication*

Urdhva-Tiryagbhyam sutra is an ancient Vedic mathematics method for multiplication [6]. It is a general formula applicable to all cases of multiplication. The formula is very short and consists of only one compound word and means 'Vertically and crosswise'. In Urdhva Tiryagbhyam algorithm, the number of steps required for multiplication can be reduced and hence the speed of multiplication is increased.

An illustration of steps for computing the product of two 4-bit numbers is shown below [9, 10]. The two input are $a_3a_2a_1a_0$ and $b_3b_2b_1b_0$ and let $p_7p_6p_5p_4p_3p_2p_1p_0$ be the product. And the temporary partial products are $t_0, t_1, t_2, \dots, t_6$. The partial products are obtained from the steps given below. The line notation of the steps is shown in fig. 5.

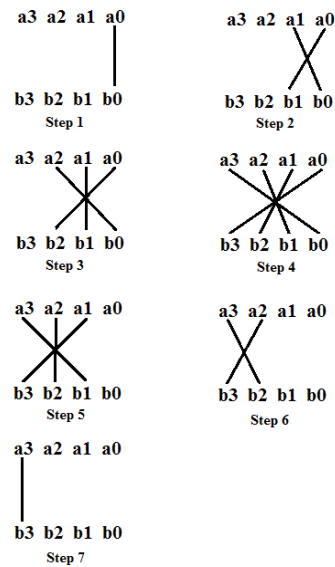

Fig. 5 Line notation of Urdhva Tiryagbhyam sutra

Step1: $t_0(1bit) = a_0b_0$.
Step2: $t_1(2bit) = a_1b_0 + a_0b_1$.
Step3: $t_2(2bit) = a_2b_0 + a_1b_1 + a_0b_2$

Step4: $t_3(3bit) = a_3b_0 + a_2b_1 + a_1b_2 + a_0b_3$.
Step5: $t_4(2bit) = a_3b_1 + a_2b_2 + a_1b_3$.
Step6: $t_5(2bit) = a_3b_2 + a_2b_3$.
Step7: $t_6(1bit) = a_3b_3$

The product is obtained by adding $s_1, s_2$ and $s_3$ as shown below, where $s_1, s_2$ and $s_3$ are the partial sum obtained.

$s_1 = t_6\ t_5[0]\ t_4[0]\ t_3[0]\ t_2[0]\ t_1[0]\ t_0$
$s_2 = t_5[1]\ t_4[1]\ t_3[1]\ t_2[1]\ t_1[1]$
$s_3 = t_3[2]$

$$
\begin{array}{c}
\text{Product} = t_6\ t_5[0]\ t_4[0]\ t_3[0]\ t_2[0]\ t_1[0]\ t_0\ + \\
t_5[1]\ t_4[1]\ t_3[1]\ t_2[1]\ t_1[1]\ 0\ + \\
t_3[2]\ 0\ 0\ 0\ 0 \\
\hline
p_7\ p_6\ p_5\ p_4\ p_3\ p_2\ p_1\ p_0
\end{array}
$$

This method can be further optimized to reduce the number of hardware. A more optimized hardware architecture [11, 12] is shown in Fig. 6. This model actually helps to eliminate the need for three operand 7-bit adder and hence reduces hardware and delay. The adders are connected in ripple manner.

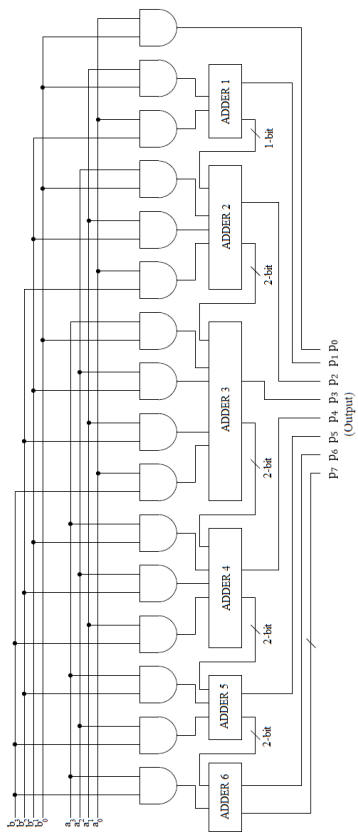

Fig. 6 Hardware architecture for 4x4 Urdhva Tiryagbhyam multiplier.

The expressions for product bits are as shown below.
$p_0 = a_0b_0$

$p_1 = LSB\ of\ (Sum(ADDER\ 1))$
   $= LSB\ of\ (a_1b_0 + a_0b_1)$
$p_2 = LSB\ of\ (Sum(ADDER\ 2))$
   $= LSB\ of\ (MSB(ADDER1) + a_2b_0 + a_1b_1 + a_0b_2)$
$p_3 = LSB\ of\ (Sum(ADDER\ 3))$
   $= LSB\ of\ (MSB(ADDER\ 2) + a_3b_0 + a_2b_1 + a_1b_2 + a_0b_3)$
$p_4 = LSB\ of\ (Sum(ADDER\ 4))$
   $= LSB\ of\ (MSB(ADDER1) + a_3b_1 + a_2b_2 + a_1b_3)$
$p_5 = LSB\ of\ (Sum(ADDER\ 5))$
   $= LSB\ of\ (MSB(ADDER1) + a_3b_2 + a_2b_3)$
$p_6 = LSB\ of\ (Sum(ADDER\ 6))$
   $= LSB\ of\ (MSB(ADDER1) + a_3b_3)$
$p_7 = Carry\ of\ ADDER$

Since there are more than two operands in adders 2 to 5, we can use carry save addition to implement adders 2 to 5. This technique reduces the delay to a great extend compared to the ripple carry adder.

*Karatsuba Algorithm for multiplication*

Karatsuba multiplication algorithm [4, 5] is best suited for multiplying very large numbers. This method is discovered by Anatoli Karatsuba in 1962. It is a divide and conquer method, in which we divide the numbers into their Most Significant half and Least Significant half and then multiplication is performed. Karatsuba algorithm reduces the number of multipliers required by replacing multiplication operations by addition operations. Additions operations are faster than multiplications and hence the speed of multiplier is increased. As the number of bits of inputs increase, Karatsuba algorithm becomes more efficient. This algorithm is optimal if width of inputs is more than 16 bits. The hardware architecture of Karatsuba algorithm is shown in Fig. 7.

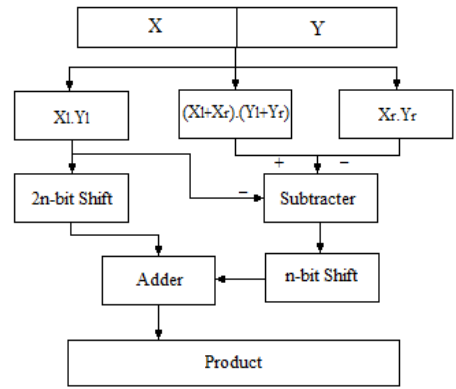

Fig. 7 Block diagram of Karatsuba multiplier

Karatsuba algorithm for two inputs X and Y can be explained as follow.
Product$= X.Y$
X and Y can be written as,
$$X = 2^{n/2}.X_l + X_r \quad (1)$$
$$Y = 2^{n/2}.Y_l + Y_r \quad (2)$$

Where $X_l$, $Y_l$ and $X_r$, $Y_r$ are the Most Significant half and Least Significant half of X and Y respectively, and n is the number of bits.
Then,

$$X.Y = \left(2^{\frac{n}{2}} X_l + X_r\right).\left(2^{\frac{n}{2}} Y_l + Y_r\right)$$
$$= 2^n. X_l Y_l + 2^{n/2} (X_l Y_r + X_r Y_l) + X_r Y_r \quad (3)$$

The Second term in equation (3) can be optimized to reduce the number of multiplication operations.
i.e.;   $X_l Y_r + X_r Y_l = (X_l + X_r)(Y_l + Y_r) - X_l Y_l - X_r Y_r$ (4)

The equation (3) can be re-written as,

$$X.Y = 2^n. X_l Y_l + X_r Y_r + 2^{\frac{n}{2}} ((X_l + X_r)(Y_l + Y_r) - X_l Y_l - X_r Y_r)$$
(5)

The recurrence of Karatsuba algorithm is,
$$T(n) = 3T\left(\frac{n}{2}\right) + O(n) \quad O(n^{1.585}) \quad (6)$$

### D. Normalization of the result

Floating point representations have a hidden bit in the mantissa, which always has a value 1 and hence it is not stored in the memory to save one bit. A leading 1 in the mantissa is considered to be the hidden bit, i.e. the 1 just immediate to the left of decimal point. Usually normalization is done by shifting, so that the MSB of mantissa becomes nonzero and in radix 2, nonzero means 1. The decimal point in the mantissa multiplication result is shifted left if the leading 1 is not at the immediate left of decimal point. And for each left shift operation of the result, the exponent value is incremented by one. This is called normalization of the result. Since the value of hidden bit is always 1, it is called 'hidden 1'.

### E. Representation of exceptions

Some of the numbers cannot be represented with a normalized significand. To represent those numbers a special code is assigned to it. In the proposed model, we use four output signals namely Zero, Infinity, NaN (Not-a-number) and Denormal to represent these exceptions. If the product has $exponent + bias = 0$ and $significand = 0$, then the result is taken as Zero (±0). If the product has $exponent + bias = 255$ and $significand = 0$, then the result is taken as Infinity (∞). If the product has $exponent + bias = 255$ and $significand \neq 0$, then the result is taken as NaN. Denormalized values or Denormals are numbers without a hidden 1 and with the smallest possible exponent. Denormals are used to represent certain small numbers that cannot be represented as normalized numbers. If the product has $exponent + bias = 0$ and $significand \neq 0$, then the result is represented as Denormal. Denaormal is represented as $\pm 0.s \times 2^{-126}$, where s is the significand.

## IV. IMPLIMENTATION AND RESULTS

The main objective of this work is to design and implement a floating point variable-precision circuit such that the device can reconfigure itself according to the precision requirements and can operate at high speed irrespective of accuracy and consume less power where accuracy is not an issue. Since mantissa multiplication is the most complex part in the floating point multiplier, we designed a multiplier which can operate at high speed and increase in delay and area is significantly less with increasing number of bits. The floating point multipliers of different modes with IEEE-754 standard format and custom precision format is implemented separately using Verilog HDL and tested. The binary multiplier unit (Karatsuba-Urdhva) are further optimized by replacing simple adders with efficient adders like carry select adders and carry save adders. The proposed model is implemented, synthesized and simulated using Xilinx Synthesis Tools (ISE 14.7) targeted on Virtex4 family. The model operates in a selected mode only and during operation, only the selected multiplier unit will be in ON state and all other multipliers units will be in OFF state. Hence, if a low precision mode is selected, the area and hence the power consumption will be less. The summary of results is given in table II and table III. Comparison with various multiplier units is given in tables IV, V, VI, VII and VIII.

TABLE II - Performance analysis of Karatsuba-Urdhva multipliers in the proposed model

|  | 8-bit multiplier | 16-bit multiplier | 24-bit multiplier | 32-bit multiplier |
|---|---|---|---|---|
| Slices | 113 | 410 | 972 | 1389 |
| LUTs | 120 | 451 | 1018 | 1545 |
| IOBs | 33 | 65 | 97 | 129 |
| Delay | 9.396ns | 11.514ns | 12.996ns | 13.141ns |
| $f_{max}$ (MHz) | 274.469 | 248.964 | 226.508 | 209.606 |
| Logic levels | 14 | 22 | 31 | 39 |

TABLE III – Performance analysis of floating point units in the proposed model

|  | 8-bit precision floating point multiplier | 16-bit precision floating point multiplier | 23-bit precision floating point multiplier | Double precision floating point multiplier |
|---|---|---|---|---|
| Slices | 157 | 475 | 977 | 3877 |
| LUTs | 220 | 584 | 1073 | 4033 |
| IOBs | 61 | 83 | 104 | 193 |
| Delay | 12.254ns | 14.577ns | 16.392ns | 18.966ns |
| $f_{max}$ (MHz) | 264.767 | 240.955 | 226.508 | 173.952 |

TABLE IV - Delay comparison of various 8-bit multipliers with proposed Karatsuba-Urdhva multiplier

|       | Ref [9] | Ref [12] | Ref [13] | Proposed multiplier |
|-------|---------|----------|----------|---------------------|
| Width | 8-bit   | 8-bit    | 8-bit    | 8-bit               |
| Delay | 28.27ns | 15.050ns | 23.973ns | 9.396ns             |

TABLE V - Delay comparison of various 16-bit multipliers with proposed Karatsuba-Urdhva multiplier

|       | Ref [14]-vedic multiplier | Ref [15] | Proposed multiplier |
|-------|---------------------------|----------|---------------------|
| Width | 16-bit                    | 16-bit   | 16-bit              |
| Delay | 13.452ns                  | 27.148ns | 11.514ns            |

TABLE VI - Delay and area comparison of 24-bit multipliers with proposed Karatsuba-Urdhva multiplier

|                     | Slices | LUTs | Delay    |
|---------------------|--------|------|----------|
| Ref [16]            | 1306   | 2329 | 16.316ns |
| Proposed multiplier | 972    | 1018 | 12.996ns |

TABLE VII - Delay and area comparison of 32-bit multipliers with proposed Karatsuba-Urdhva multiplier

|                                           | LUTs | Delay    |
|-------------------------------------------|------|----------|
| Ref [14]- Modified Booth multiplier (Radix-8)  | 2721 | 12.081ns |
| Ref [14]- Modified Booth multiplier (Radix-16) | 7161 | 11.564ns |
| Ref [14]                                  | 2704 | 9.536ns  |
| Proposed multiplier                       | 1545 | 13.141ns |

TABLE VIII - Delay and area comparison of SP-floating point multiplier with proposed SP FP multiplier

|                     | Slices | LUTs | Delay    |
|---------------------|--------|------|----------|
| Ref [16]            | 1269   | 2270 | 18.783ns |
| Ref [8]             | 1149   | 1146 | --       |
| Proposed multiplier | 976    | 1091 | 16.392ns |

## V. CONCLUSION AND FUTURE WORK

This paper describes a method to effectively adjust the delay and power consumption for different accuracy requirements. Also the paper shows how to effectively reduce the percentage increase in delay and area of a floating point multiplier with increase in number of bits by using a very efficient combination of Karatsuba and Urdhva-Tiryagbhyam algorithms. The model can be further optimized in terms of delay by using pipelining methods and precision of the result can be increased by adding efficient truncation and rounding methods.